\documentstyle[12pt,psfig]{article}

\begin{document}
\input epsf

\title{{\bf The Phase Transition to a Square Vortex Lattice in
Type-II Superconductors with Fourfold Anisotropy}}
\author{Kyungwha Park and  David A. Huse \\
{\small \it Department of Physics, Princeton University, Princeton, NJ 08544} }
\date{{\small \today}}
\maketitle

\begin{abstract}
We investigate the stability of the square vortex lattice which
has been recently observed in experiments on the
borocarbide family of superconductors.
Taking into account the tetragonal symmetry of these systems, we add
fourfold symmetric fourth-derivative terms to the
Ginzburg-Landau(GL) free energy.  At $H_{c2}$ these terms may be treated
perturbatively to lowest order to locate the 
transition from a distorted hexagonal to a square vortex lattice.
We also solve for this phase boundary numerically
in the strongly type-II limit, finding
large corrections to the lowest-order perturbative results.
We calculate the relative fourfold $H_{c2}$ anisotropy
for field in the $xy$ plane
to be 4.5\% at the temperature, $T_c^{\Box}$, where the transition occurs
at $H_{c2}$ for field along the $z$ axis.  This is to be compared
to the 3.6\% obtained in the perturbative calculation.
Furthermore, we find that the phase boundary in the $H-T$ phase diagram
has positive slope near $H_{c2}$.
\end{abstract}


\section{Introduction}
The possibility of square vortex lattices in type-II
superconductors was first proposed by
Abrikosov \cite{abrikosov} in his original paper on
vortex lattices.  However, it was later shown that the
hexagonal vortex lattice has a lower free energy for
isotropic materials. \cite{kra}
Anisotropy in the effective mass or superfluid density
tensors cause distortions of this hexagonal lattice
simply in proportion to the anisotropy of the
magnetic penetration length.  Effects due to higher-order
anisotropies (e.g., in cubic crystals) were seen
in the 1970's for some low $\kappa$ 
superconductors such as PbTl, PbBi, and Nb. \cite{weber}
In some cases, when the external magnetic field was applied
along a fourfold symmetric axis of the
crystal, the vortices formed a square lattice.  When the field was applied
along the threefold or twofold symmetric axis of the crystals,
hexagonal or distorted hexagonal lattices were observed, respectively. 
In the case of Nb, for field along [001] 
the square vortex lattice transformed into a
hexagonal lattice as
the temperature or field increased.
Takanaka \cite{takanaka} introduced terms with cubic anisotropy
into the phenomenological Ginzburg-Landau(GL) theory to
explain the vortex lattice structures quite successfully. 
The extra terms arise from
anisotropy of the Fermi surface.

Recently, in ${\rm ErNi_2B_2C}$, one of the borocarbide family
of superconductors, 
Yaron, {\it et al.}\cite{yaron} and Eskildsen, {\it et al.}\cite{gammel} have
used neutron scattering and Bitter decoration to observe the phase 
transition from a distorted hexagonal vortex lattice aligned
with either the [100] or [010] directions to a square vortex lattice
aligned with the [110] direction 
for increasing magnetic field along the [001] direction.
This was the first observation of a square vortex
lattice in a strongly type-II material with $\kappa>>1/\sqrt{2}$.
These experiments have motivated further
theoretical and experimental studies. 
Kogan {\it et al.} \cite{kogan} have analyzed the vortex lattice transition 
in borocarbides 
by taking into account the fourfold anisotropic corrections to the
London model in the intermediate field region and $T \ll T_{c}$.
This study revealed two kinds of phase transitions:
at small fields (along the [001] direction), the distorted hexagonal lattice
is aligned with a [110] direction. As the field increases, the lattice
reorients along a [100] direction in a first-order phase transition. 
At a still higher field the square vortex lattice becomes stable.  
The latter phase transition is continuous.  The square vortex lattice 
has also been observed in other borocarbides such as
${\rm LuNi_2B_2C}$ \cite{wilde} and 
${\rm YNi_2B_2C}$ \cite{yeth}, using scanning tunnelling
microscopy (STM) and neutron scattering, respectively. 

Vortex lattice structures in the high-temperature superconductors
have also been studied.  For low applied fields along the $c$ axis
(normal to the copper-oxide planes), the flux lattice can be
seen using Bitter decoration.  A distorted hexagonal lattice
was seen for YBCO, \cite{dolan} distorted as expected from the
in-plane ($ab$) effective mass anisotropy.  But the STM pictures
of the vortex lattice obtained at much higher field \cite{ma} are
more consistent with a square lattice that is distorted by
effective mass anisotropy.  Thus it appears that the hexagonal
to square lattice transition that we are studying may also
occur in YBCO.  Motivated by this, Affleck {\it et al.} \cite{affleck} 
showed how fourfold anisotropic terms can arise in the London
free energy due to a
$d$-wave symmetry of the superconducting order parameter.

In this paper, we examine the structure of the vortex lattice
near $H_{c2}$ within a generalized Ginzburg-Landau (GL) free energy
functional.  We consider systems with tetragonal symmetry such as the
borocarbides.  The lowest-order term (ignoring
effective mass anisotrpy) that reduces the symmetry
from isotropic down to tetragonal is fourth-order in the 
spatial derivative and second-order in the order parameter.
We discuss the three distinct terms that can be added to the GL
free energy at this order.  These new terms may be treated
perturbatively near $H_{c2}$ to first order, demonstrating that
the square lattice occurs at high fields where the higher-order
gradient terms are more important, while the lattice distorts toward
hexagonal at lower fields, in qualitative agreement with experiment.  
The phase transition, which we treat only on a mean-field
level, is an Ising-type critical point, with the
square unit cell distorting into a rhombus in
two possible ways by increasing the length of one
of its diagonals and reducing the other. 
The perturbation expansion has not been carried 
to higher order, so the quantitative
accuracy of the first-order approximation had not previously been estimated.
To investigate this we have numerically determined the phase diagram
of this system, locating the upper critical field for various field 
orientations and the phase boundary separating the square and
distorted hexagonal vortex lattice phases.  We find that there
are substantial (up to 50\%) differences from the perturbative
approximation, for example in the location of the phase boundary.
Thus although the first-order perturbative approximation gives
a good qualitative description of the phase diagram, quantitatively
accurate results require a proper treatment of higher-order effects. 

\section{Ginzburg-Landau model}
To study the lowest free energy pattern of the vortex lattice
near $H_{c2}$ for tetragonal crystals, we use
the Ginzburg-Landau (GL) theory
with added fourfold symmetric fourth-order derivative terms.
The full free energy functional that we minimize
(ignoring thermal fluctuations) is
\begin{eqnarray}
F & = & F_{0} + F_{4} . \label{eq:f-GL} 
\end{eqnarray}
$F_0$ is the usual GL free energy:
\begin{eqnarray}
F_{0} & = & \int d^3r \{ \alpha |\psi|^2 + \frac{\beta}{2}
|\psi|^4 + \frac{1}{2m^{\ast}} |\vec{\Pi} \psi|^2
+ \frac{h^2}{8 \pi} \}   .
\end{eqnarray}
For simplicity, we have taken an isotropic effective mass tensor.
Tetragonal symmetry, of course, also allows uniaxial effective mass
anisotropy, but this can be scaled out by rescaling the $z$ axis
and the $xy$ components of the magnetic field.  The only resulting
difference is then in the $h^2$ field energy term in $F_0$, but
this does not affect any of our results below.
$F_4$ contains the additional terms allowed with tetragonal symmetry
that are of fourth order in the $x$ and $y$ 
gauge-invariant derivatives
and second order in the order parameter: 
\begin{eqnarray}
F_{4} & = & \int d^3r \frac{1}{4m^{\ast 2}|\alpha|}[\epsilon_i \tau
|(\Pi_x^2 + \Pi_y^2) \psi|^2 \nonumber \\
& &+ \epsilon_a \tau
(|(\Pi_x^2 - \Pi_y^2)\psi|^2
- |(\Pi_x \Pi_y + \Pi_y \Pi_x)\psi|^2) \nonumber \\
& &+ \frac{\epsilon_b \tau \hbar^2 e^{\ast 2}}{c^2} 
h_z^2 |\psi|^2]  .
\end{eqnarray}
The gauge-invariant derivative is
\begin{eqnarray}
\vec{\Pi}&=& \frac{\hbar}{\imath}\vec{\nabla} - \frac{e^{\ast}}{c}\vec{A}  . 
\end{eqnarray}
We are interested in $T < T_c$, so $\alpha$ is negative. 
$\tau$ is defined to be $1 - \frac{T}{T_c}$, 
so it is positive. $\vec{h}(\vec{r})$ is the local
magnetic field, $\vec{h}=\vec{\nabla} \times \vec{A}$.  
$F_{4}$ contains
two isotropic terms with coefficients $\epsilon_i$ and 
$\epsilon_b$, and the anisotropic term with coefficient
$\epsilon_a$.  The approximation of limiting the free energy to
include only terms that are low-order in the gradient and the
order parameter is valid for temperatures near $T_c$.  Only terms that
are of fourth or higher order in the $x$ and $y$ gradients can reduce the
rotational symmetry about the $z$-axis from circular down to
tetragonal, which is why we add the fourth-order terms to the
usual GL free energy.  The anisotropic term may arise simply from fermi-surface
anisotropy \cite{takanaka} or from a d-wave nature of the pairing
\cite{affleck}.

In the literature \cite{affleck,wilde},
certain particular combinations of the isotropic and the 
anisotropic fourth order derivative terms in $F_{4}$ have been used.
In our numerics (below), we will also use a specific 
convenient combination of them.  For the theory as
specified above to
be stable, the fourth-derivative terms have to be
non-negative, which requires $\epsilon_i \geq |\epsilon_a|$.
In our notation, a term $\epsilon \tau |(\Pi_x^2 -\Pi_y^2) \psi|^2$ \cite{affleck}
corresponds to
\begin{eqnarray}
\epsilon_i,\epsilon_a,\epsilon_b&=&\frac{\epsilon}{2},
\frac{\epsilon}{2},\frac{3 \epsilon}{2} .
\end{eqnarray}
A term $\epsilon \tau (|\Pi_x^2 \psi|^2 + |\Pi_y^2 \psi|^2)$ \cite{wilde} 
corresponds to
\begin{eqnarray}
\epsilon_i,\epsilon_a,\epsilon_b&=&\frac{3\epsilon}{4},
\frac{\epsilon}{4},\frac{3 \epsilon}{4} .
\end{eqnarray}
And the term $\epsilon \tau |(\Pi_x \Pi_y + \Pi_y \Pi_x)\psi|^2$ that we use in
our numerics corresponds to 
\begin{eqnarray}
\epsilon_i,\epsilon_a,\epsilon_b&=&\frac{\epsilon}{2},
-\frac{\epsilon}{2},\frac{3 \epsilon}{2} .
\end{eqnarray}
In $F_{4}$, the anisotropic term is the one that 
causes the vortex lattice to distort and become square
when the field is along the $z$ axis and $|\epsilon_a|\tau$
is large enough.  The isotropic terms do not by
themselves cause any distortion of the hexagonal vortex lattice,
but they do change $H_{c2}$.

For convenience, we introduce the following ``dimensionless'' units:
the magnetic field is measured in units of $H_{c2}=\phi_0/(2\pi\xi^2)$,
the vector potential in units of $H_{c2} \xi$,
the length in units of the coherence 
length $\xi=\hbar/ \sqrt{2m^{*}|\alpha|}$, the order parameter
in units of the zero-field order parameter magnitude:
$|\psi_{\infty}|=\sqrt{|\alpha|/\beta}$, and the energy in units
of $H_c^2 \xi^{3} / 4 \pi$.
Using these units, the 
free energy (\ref{eq:f-GL}) is
\begin{eqnarray}
F &=& \int d^3r [-|\psi|^2 + \frac{1}{2} |\psi|^4 +
|\vec{\Pi} \psi |^2 + \kappa^2 h^2 +
\epsilon_i \tau |(\Pi_x^2 + \Pi_y^2) \psi|^2 \nonumber \\
& & + \epsilon_a \tau
(|(\Pi_x^2 - \Pi_y^2)\psi|^2
- |(\Pi_x \Pi_y + \Pi_y \Pi_x)\psi|^2)  \nonumber \\
& &+ \epsilon_b \tau h_z^2 |\psi|^2]    ,
\end{eqnarray}
with
\begin{eqnarray}
\vec{\Pi}&=&\frac{\vec{\nabla}}{\imath} - \vec{A}  . 
\end{eqnarray}

\section{Perturbative results}
First we look at the vortex lattice behavior at
$H_{c2}$ by treating $F_{4}$ as a perturbation to the
usual GL theory, $F_{0}$, obtaining the free
energy to first order in $F_{4}$.
Most of these results were first presented
by Takanaka \cite{takanaka}, and then, 
much more recently, by De Wilde, {\it et al.} \cite{wilde} .

Because of $F_{4}$, the upper critical
field depends on the field orientation.
When the field is along the $z$ axis, the upper critical field
(in the absence of fluctuations) is given by
\begin{eqnarray}
H_{c2}[001] &=& \frac{\phi_0}{2 \pi \xi^2} ( 1 -
\epsilon_i \tau -\epsilon_b \tau + ...)  . 
\end{eqnarray}
When the field is in the $xy$ plane, 
\begin{eqnarray}
H_{c2}[xy0] &=& \frac{\phi_0}{2 \pi \xi^2} 
(1 - \frac{3}{4} ( \epsilon_i \tau + \epsilon_a \tau \cos 4\phi) 
+ ...),
\end{eqnarray}
where $\phi$ is the angle between the field and the $x$ axis.
The relative $xy$-plane anisotropy of the upper critical field
is thus
\begin{eqnarray}
\Delta H_{c2}/H_{c2}&=&(H_{c2}[100] - H_{c2}[110])/H_{c2}[100] \\
                    &=&-\frac{3\epsilon_a \tau}{2} + ....
\end{eqnarray}

Near $H_{c2}$ the GL free energy density, including the first-order
perturbative contributions from $F_4$, may be simplified
by the Abrikosov identities\cite{abrikosov,takanaka,super}.
For the field along the $z$ axis, this gives
\begin{eqnarray}
f&=&\kappa^2[B^2-\frac{(B-H_{c2}(\epsilon_i \tau,\epsilon_b \tau))^2}
{1-\frac{<h_s^2>}{<h_s>^2}+ \frac{<|\psi_L|^4>}
{2 \kappa^2 <h_s>^2} } ]  ,
\label{eq:gen-energy}
\end{eqnarray}
where $\psi_L$ is the solution of the linearized GL equation 
at $H_{c2}$, including the terms from $F_4$.
$h_s$ is defined by $h = H + h_s$, where
$H$ is the applied field; all fields are along the $z$ axis.
$< \cdots >$ means a spatial average over the vortex lattice,
and $B = <h>$ is the spatial average of the magnetic field.
The free energy density (\ref{eq:gen-energy}) can be written
in terms of the Abrikosov ratio $\beta_A$ as 
\begin{eqnarray}
f &=& \kappa^2 [ B^2 -
\frac{(B-H_{c2}(\epsilon_i \tau, \epsilon_b \tau))^2}
{1+(2\kappa^2-1) \gamma \beta_A(
\epsilon_a \tau)} ]  , \\
\gamma&\equiv&1- \frac{8 \kappa^2(\epsilon_i \tau + \epsilon_b \tau)
-\frac{2 \epsilon_b \tau}{3}}{2\kappa^2 -1}  , \nonumber
\end{eqnarray}
where 
\begin{eqnarray}
\beta_A &\equiv& \frac{<|\psi_L|^4>}{<|\psi_L|^2>^2} \nonumber \\
        &=& \beta_0
- \epsilon_a \tau Re \beta_4  ,
\end{eqnarray}
\begin{equation}
\beta_{2n} = \frac{<(\Pi_{+}^{2n} \psi_0 ) \psi_{0}^{\ast}
|\psi_{0}|^2> }{<|\psi_0|^2>^2 2^{n} }  .
\end{equation}
$\psi_0$ is the solution to the linearized GL equations 
when $\epsilon_{i,a,b}=0$,
and $\Pi_{+}$ is $\Pi_x - \imath \Pi_y$.
We obtain the stable pattern of the vortex lattice 
just below $H_{c2}$ by minimizing
the Abrikosov ratio, $\beta_A$.
For regular vortex lattices with one vortex per unit
cell, the lattice vectors are specified by (see Fig. 1)
\begin{eqnarray*}
\vec{a_1} & = & (\frac{L_x}{2}, - \frac{L_y}{2}) ,\\
\vec{a_2} & = & (\frac{L_x}{2}, \frac{L_y}{2}).
\end{eqnarray*}
(The vortex lattice may also be rotated from this orientation
by angle $\pi/4$; this is equivalent to changing the sign of $\epsilon_a$.)
$\beta_0$ and $\beta_4$ can be written explicitly using the reciprocal lattice
vectors as
\begin{eqnarray}
\beta_{0} & = & \sum_{p,q} \exp{ [ - \frac{\pi}{2b} \{
(p^2 + q^2)(1 + b^2) + 2pq (1 - b^2) \} ] } , \\
\beta_{4} & = & \sum_{p,q} \frac{\pi^2}{4b^2} (p + q + \imath
(-p + q)b )^4  \nonumber  \\
          &   & \exp{ [ - \frac{\pi}{2b} \{ (p^2 + q^2)(1 + b^2) +
2pq(1 - b^2) \} ]  } ,
\end{eqnarray}
where $b = L_x / L_y$. 
The sums run over all (positive, zero and negative) integer pairs.
The square vortex lattice is $b=1$ and the hexagonal lattice is
$b=\sqrt{3}$ or $b=1/\sqrt{3}$; other values of $b$ we call
distorted hexagonal lattices.

For small positive $|\epsilon_a|\tau$ the lowest free energy states
for this system are distorted hexagonal lattices,
and the square vortex lattices
are local maxima of the free energy.  
To calculate the stability of the square lattice
the second derivatives at $b=1$ are needed: they are
$d^2\beta_{0}/db^2 \cong -0.295$ and
$d^2 Re \beta_{4}/db^2 \cong 12.4$.
Thus, within the approximation of calculating $\beta_A$ to only first order
in $\epsilon_a$, at $H_{c2}$ the appropriately-oriented square vortex lattice
changes from a local maximum of the free energy to a global
minimum at 
\begin{eqnarray}
|\epsilon_a|\tau_{c}^{\Box} \cong 0.0238  .
\end{eqnarray}
For larger values of $|\epsilon_a|\tau$ the lowest
free energy state of the vortex lattice is square.
This critical value of $|\epsilon_a|\tau$ does not depend on
$\kappa$ at first order in $\epsilon_a$.
This first-order approximation then says that
the phase transition from distorted hexagonal to square vortex
lattice at $H_{c2}$ for the field along the $z$ axis is at the
temperature where the $xy$-plane $H_{c2}$ varies by
about 3.6\% from [100] to [110] field directions.

One way to get a sense of how accurate this first-order in
$F_{4}$ approximation is would be to calculate the next
corrections (second-order in $F_{4}$).  This did not appear
to be analytically tractable to us, so we have instead 
located the phase transition
numerically, finding that there are indeed rather large
corrections to this first-order approximation.

\section{Numerical Study}

At the order we are considering, in $F_4$ there are three different
terms added to the usual Ginzburg-Landau theory, and $F_0$ contains
$\kappa$ and $\vec{H}$, which cannot be scaled away.   In
principle, one could examine the behavior in the full
parameter space $(\vec{H},\kappa,\epsilon_i,\epsilon_a,\epsilon_b)$.
However, we are interested in the phase transition to the
square vortex lattice that is caused by the anisotropic
term, $\epsilon_a$.  If we add only this term, 
it can be negative so the free energy then becomes 
unbounded below and the system is unstable at short
wavelengths.  To stabilize the system we set 
$\epsilon_i = |\epsilon_a|$.  We also set $\epsilon_b$
to keep $H_{c2}$ at its original value of 
$\phi_0/(2\pi\xi^2)$ for field along the $z$ axis.  For simplicity, we will
only examine the large $\kappa$ limit, where the magnetic field
is uniform.  With all these constraints and simplifications, the
system we studied numerically is, in dimensionless
form:
\begin{eqnarray}
F&=&\int d^3 r [  -|\psi|^2 + \frac{1}{2} |\psi|^4 +
|\vec{\Pi} \psi |^2 +  \nonumber \\
 &  & \epsilon\tau|(\Pi_x\Pi_y+\Pi_y\Pi_x)\psi|^2  
 +  \epsilon_b \tau H_z^2 |\psi|^2]  ,
\label{eq:GL}
\end{eqnarray}
with $\epsilon\tau = 2 \epsilon_i\tau = -2 \epsilon_a\tau \geq 0$  
and for each value of $\epsilon\tau$, $\epsilon_b\tau$ is chosen so
that along the $z$ axis
$H_{c2} = 1$ in these units.  In the first-order
approximation above, for field along the $z$ axis
the phase transition to the square vortex
lattice at $H_{c2}$ occurs at $\epsilon\tau \cong 0.048$.  
We find the higher-order
corrections are substantial, and the actual transition 
in this system (\ref{eq:GL}) does
not occur until $\epsilon \tau_c^{\Box} \cong 0.073$.  
Presumably the precise result for $\epsilon_a\tau_c^{\Box}$ 
will vary, depending on the values of $\kappa,\epsilon_i$
and $\epsilon_b$ as well as other higher-order terms, all of
which potentially differ between materials; 
we have not explored this dependence.  Our basic message here is
that the first-order in $\epsilon_a$ perturbative estimate is
substantially off, so higher-order effects have to be treated properly
to quantitatively estimate the phase diagram for any particular system. 

The stable vortex lattice for this system (\ref{eq:GL})
with the field along the $z$ axis is shown in Fig. 1.  
The area of the unit cell is dictated by the magnetic
field because there is exactly one flux quantum per vortex.  The only remaining
freedom is the angle, $\theta$, between the two nearest-neighbor
vectors shown in Fig. 1.  
The square lattice is $\theta = \frac{\pi}{2}$,
the hexagonal lattice is $\theta = \frac{\pi}{3}$ (or $\frac{2\pi}{3}$);
other angles are the distorted hexagonal lattices.
The minimum free energy vortex lattice for each $\theta$ is an
order parameter and current pattern that is symmetric under the
operations that reverse the currents and then reflect
in a plane that contains vortex lines and is parallel to the $x$
or $y$ axis.  These symmetries
mean we only have to determine the pattern over 
one quarter of the unit cell of the vortex lattice.
We have chosen the quarter in the positive $x$ and $y$ quadrant
(see Fig. 1).

We discretize the system on a rectangular numerical grid 
aligned with the $x$ and $y$ axes.   We use
gauge-invariant variables only, namely the magnitude $|\psi|$
of the order parameter at each grid point, and the gauge-invariant 
phase differences between nearest neighbor grid points
along the $x$ direction, $\Delta_x\phi$, and along the
$y$ direction, $\Delta_y\phi$.  These phase differences
are not all independent, since their sum around an 
elementary plaquette of the grid must be equal to $Hd_x d_y$,
where $d_x$ and $d_y$ are the grid spacings.
We put one grid point at the
precise core of the vortex at the origin in Fig. 1; at this point
$|\psi| = 0$ so the phase is ill-defined and special care
must be taken when treating the terms involving this point.

All terms in the discretized GL free energy are
symmetric so that the differences between the 
discretized expression and the continuum limit are
even polynomials of $d_x$ and $d_y$.  We extrapolate to the
continuum limit by reducing $d_x$ and $d_y$ together, keeping
their ratio constant.  Under these conditions
the discretization errors are proportional to $d_xd_y$, the area of the
grid plaquette, for small $d_x$ and $d_y$.
For example, under our discretization
\begin{eqnarray}
|(\frac{\nabla_x}{\imath}-A_x)\psi(x,y)|^2&\rightarrow& \nonumber \\
 \frac{1}{d_x^2}|[|\psi(x+\frac{d_x}{2},y)|\exp({\frac{\imath \Delta_x \phi}{2}})
&-&|\psi(x-\frac{d_x}{2},y)|\exp({-\frac{\imath \Delta_x \phi}{2}})]|^2  .
\end{eqnarray}
The boundary conditions on our quarter unit cell are
\begin{eqnarray}
|\psi(0,0)|&=&0,  \\
|\psi(x,\frac{L_y}{4})|&=&|\psi(\frac{L_x}{2}-x,\frac{L_y}{4})|,  \\
\Delta_x\phi(x,0)&=&0,  \\
\Delta_y\phi(0,y)&=&0,  \\
\Delta_y\phi(\frac{L_x}{2},y)&=&0,  \\
\Delta_x\phi(x,\frac{L_y}{4})&=&-\Delta_x\phi(\frac{L_x}{2}-x,\frac{L_y}{4}),\\ 
\Delta_y\phi(x,\frac{L_y}{4})&=&-\Delta_y\phi(\frac{L_x}{2}-x,\frac{L_y}{4})  ;
\end{eqnarray}
we put grid points along all these boundaries.

We begin with 
smooth initial conditions that satisfy all of the above constraints.
Then we minimize the discretized version of the GL free energy (\ref{eq:GL})
using the relaxation method \cite{adler}
\begin{equation}
\psi^{new} = \psi^{old} - \Gamma \frac{\partial F}
{\partial \psi^{\ast}}|_{old}  ,
\end{equation}
where $\Gamma$ is a relaxation parameter which can be adjusted
to avoid numerical instabilities and 
to hasten the convergence to the minimum. 

We have obtained the minimum of the discretized free energy for grid sizes
\((12\times 6),(16\times 8),(20\times 10),(24\times 12),
(28\times 14),{\rm and} (32\times 16) \) points
for the square ($\theta=90^{\circ}$) and two weakly distorted vortex lattices
($\theta=88^{\circ},86^{\circ}$).  For each $H$ and $\epsilon\tau$ the
dependence of the free energy on $\theta$ is
extrapolated to the continuum limit.
At $H_{c2}$ we also obtain the Abrikosov ratio $\beta_A(\epsilon\tau, \theta)$.
The critical parameter value above which the minimum Abrikosov ratio is at
the square lattice ($\theta=90^{\circ}$) is $\epsilon\tau_c^{\Box} \cong 0.0734$. 
To determine the phase boundary below $H_{c2}$, we fix $\epsilon\tau$ and 
sweep the magnetic field to find the transition field, $H^{\Box}$,
above which the square vortex lattice is stable.
Table \ref{table:transition} shows the transition fields 
for a few $\epsilon\tau$ values.
\begin{table}
\begin{center}
\caption{Transition fields, $H^{\Box}$, for a few $\epsilon\tau$ values.}
\label{table:transition}
\begin{tabular}{|l|l|} \hline \hline
$\epsilon\tau$ & $H^{\Box}/H_{c2}$  \\ \hline
0.0734 & 1.000 \\
0.0800 & 0.901 $\pm$ 0.002 \\
0.0860 & 0.826 $\pm$ 0.001 \\
0.0930 & 0.752 $\pm$ 0.001 \\
0.1000 & 0.687 $\pm$ 0.001 \\ \hline \hline
\end{tabular}
\end{center}
\end{table}
Within our model, we certainly could follow the phase boundary
to lower fields and temperatures, deeper into the vortex
lattice phase.  But the model, in the usual approach
of Landau theory, has kept only certain 
terms that are low-order in an expansion in the order parameter, 
the gradients, and $\tau$.  The neglected higher-order terms
become more important as one moves away from $T_c$ to lower temperature 
and away from $H_{c2}$ to lower fields.  Thus the phase diagrams of
real systems with different higher-order terms will increasingly diverge
from the behavior we find in this specific model (\ref{eq:GL}) 
as one goes to lower temperature.
This is why we did not bother to
follow the phase boundary to lower temperature
than shown in Table \ref{table:transition}.

The physically meaningful phase diagram in the magnetic field,
temperature plane is shown in Figure \ref{figure:phase}, with the
field and temperature axes scaled by the values of $H$ and $\tau$
at the point $(H_c^{\Box},\tau_c^{\Box})$
where the phase boundaries $H^{\Box}(T)$ and 
$H_{c2}(T)$ meet.  Although the point at
which this happens is at roughly 50\% larger $\epsilon\tau$ than
the estimate from the perturbative results, our phase diagram when
scaled this way is very close to the one obtained perturbatively
by De Wilde, {\it et al.},\cite{wilde} with the distorted
hexagonal/square phase boundary meeting the upper critical field
with a positive slope in the $H,T$ plane (with this scaling, the
slope of $H^{\Box}(T)$ is about 0.2, while the slope of
$H_{c2}(T)$ is -1.)  

Since, for field along the $z$ axis,
the critical $\epsilon\tau_c^{\Box}$ at $H_{c2}$ obtained numerically
is quite different from the
perturbative estimate, we have also 
numerically calculated for this system (\ref{eq:GL}) the relative
anisotropy of $H_{c2}$ in the $x$-$y$ plane at $\epsilon\tau_c^{\Box}$.
We find that the dependence
of $H_{c2}$ on the angle $\phi$ between the field and the $x$ axis
contains substantial harmonic content beyond the basic $\cos(4\phi)$
term that is present in the first-order estimate, again indicating
contributions beyond first-order in $\epsilon$. At 
$\epsilon\tau = \epsilon\tau_c^{\Box}$, the
in-plane $H_{c2}$ variation from [100] to [110] directions
is $\Delta H_{c2}/H_{c2} \cong 4.5\%$, which is again 
larger than the 3.6\% estimated perturbatively.  The in-plane
anisotropy of $H_{c2}(\phi)$ has recently been measured for
LuNi$_2$B$_2$C, where it grows to near 10\% at low temperature,
indicating that the transition to the square lattice does
meet the upper critical field for this material.\cite{metl}
 
\section{Conclusion}
We have numerically determined the vortex lattice phase diagram
for a generalized Ginzburg-Landau model (\ref{eq:GL}) that includes
higher-order terms that reduce the rotational symmetry to
tetragonal.  Although the results are
qualitatively unchanged from the estimates obtained by treating
the anisotropic term perturbatively to only first order, there
are substantial quantitative differences.  In particular, we
find that for field along the $z$ axis, the phase boundary
in the field-temperature plane
separating the distorted hexagonal and square vortex lattices
meets the upper critical field line with a positive slope at a
temperature $T_c^{\Box}$.  If the field is instead put in the
$xy$ plane at this same temperature, the difference in $H_{c2}$
between the [100] and [110] directions is about 4.5\%.  Although
this number must be a better estimate than the 3.6\% obtained
perturbatively, the full Ginzburg-Landau theory for any given
material will contain other higher-order terms that may also
have a small influence on this critical anisotropy value.  
Thus the precise value may vary somewhat away from our estimate.

\pagebreak

\begin{figure}
\begin{center}
\leavevmode
\epsfysize=8cm
\epsfbox{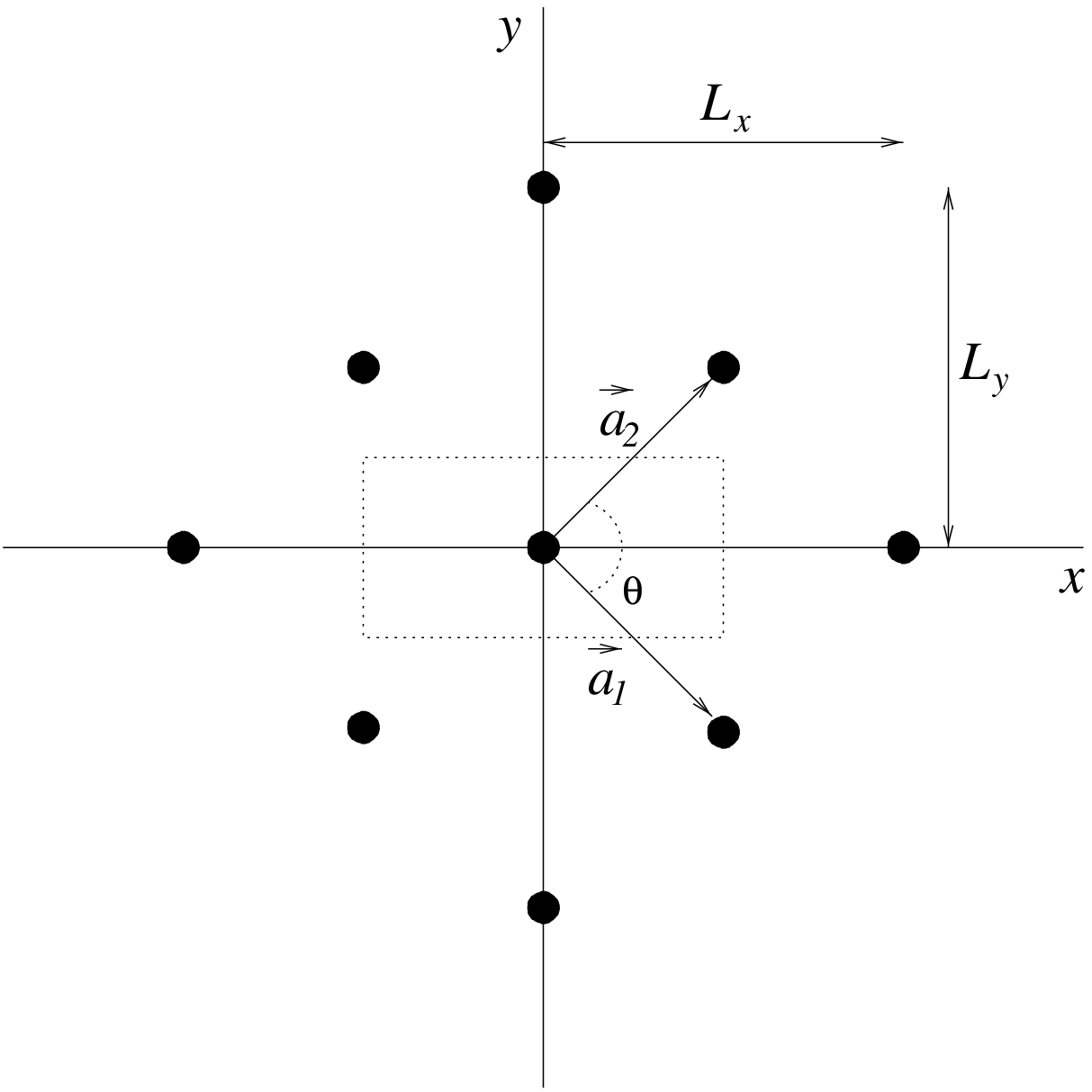}
\caption{Geometry of the vortex lattice.
The unit cell is drawn by the dotted line. The black dots
represent vortices. $\vec{a}_1$, $\vec{a}_2$ are the
lattice vectors.}
\label{geo}
\end{center}
\end{figure}
\pagebreak

\begin{figure}
\begin{center}
\leavevmode
\epsfysize=12cm
\epsfbox{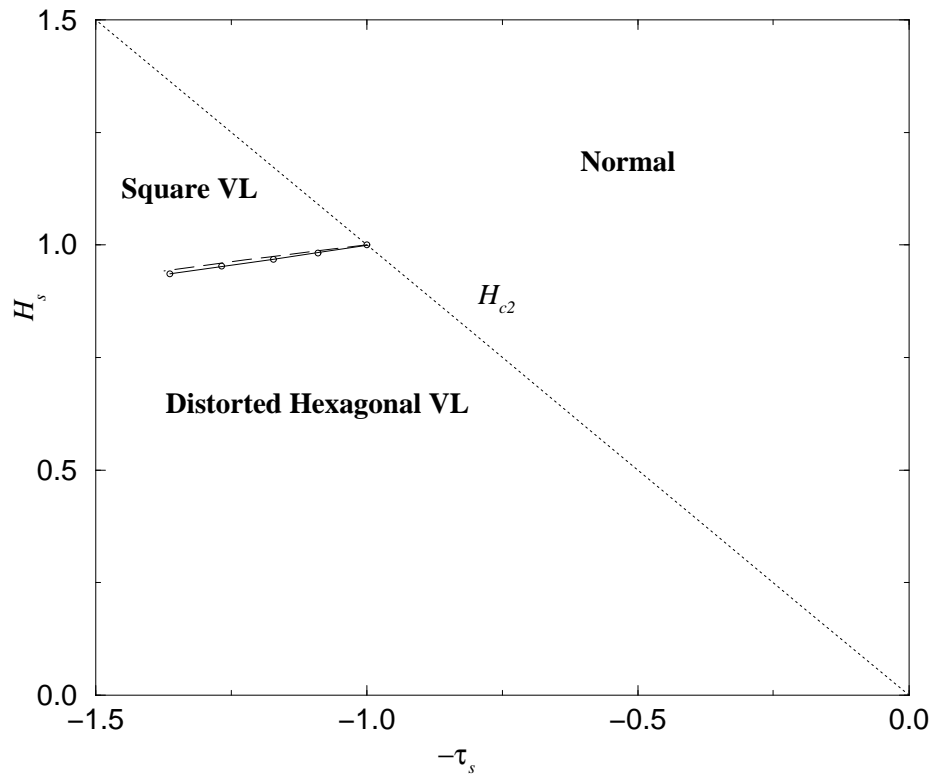}
\caption{Phase diagram of our generalized Ginzburg-Landau 
model (\ref{eq:GL}), scaled to the multicritical points
$(T_c,H=0)$ and $(T_c^{\Box},H^{\Box}_{c})$. 
The horizontal axis is the scaled
$T-T_c$: $-\tau_s = -\tau / \tau^{\Box}_{c}$,
and the vertical axis is the scaled
field: $H_s = H/H^{\Box}_{c}$.
The open circles represent our data points   
fitted to a polynomial function (solid line).
For comparison, De Wilde,{\it et al.}'s result, 
scaled the same way, is drawn by the dashed line. 
}
\label{figure:phase}
\end{center}
\end{figure}

\end{document}